\documentclass{aa}

\usepackage{graphicx}

\usepackage{natbib} \bibpunct{(}{)}{;}{a}{}{,}

\begin{document}

\title{On the non-thermal high energy radiation of galaxy clusters}

\author{%
  A. N.  Timokhin\inst{1}, F. A.  Aharonian\inst{2}, A. Yu.
  Neronov\inst{3} }

\offprints{F. A. Aharonian, \\
  \email{Felix.Aharonian@mpi-hd.mpg.de} }

\institute{Sternberg Astronomical Institute, Universitetskij pr. 13,
  Moscow, 119992, Russia \and Max-Planck-Institut f\"ur Kernphysik,
  Saupfercheckweg 1, Heidelberg, 69117, Germany \and Institute de
  Physique Theorique, Universite de Lausanne, BSP 1015 Lausanne,
  Switzerland }

\authorrunning{A.~N. Timokhin et al.}

\titlerunning{Non-thermal X-radiation from galaxy clusters}

\date{Received / Accepted}

\abstract{%
  The origin of the nonthermal EUV and hard X-ray emission ``excess''
  reported from some galaxy clusters has been intensively debated over
  last several years.  The most favored models which refer this excess
  to relativistic electrons upscattering the 2.7 K CMBR generally
  requires very low magnetic field, significantly below the estimates
  derived from the Faraday Rotation Measurements, unless one invokes
  rather nonstandard assumptions concerning the energy and pitch angle
  distributions of nonthermal electrons.
  In this paper we suggest a new model assuming that the
  ``nonthermal'' excess is due to synchrotron radiation of
  ultrarelativistic (multi-TeV) electrons of ``photonic'' origin.
  These electrons are continuously introduced throughout the entire
  intracluster medium by very high energy (hypothetical) $\gamma$-rays
  through interactions with the diffuse extragalactic radiation
  fields.  We present numerical calculations for the Coma cluster, and
  briefly discuss implications of the model for other galaxy clusters
  both in the X- and $\gamma$-ray energy domains.   }

\maketitle

\section{Introduction}

Galaxy clusters -- the largest gravitationally bound structures in the
Universe -- contain many hundreds or thousands member galaxies
surrounded by diffuse hot gas with temperature close to $\sim 10^8$K.
Correspondingly, these objects are characterized by intensive thermal
X-ray emission. At the same time, nonthermal processes connected with
acceleration and radiation of relativistic particles play
non-negligible role in the energy budget of these objects.  In
particular, diffuse radio emission with steep energy spectrum and low
surface brightness is observed from many galaxy clusters
\citep{giovannini/::radio_clusters:1999}.
Polarization measurements provide strong evidence for the synchrotron
origin of this emission.  The recent reports of the  extreme-ultraviolet
(EUV) ``excess'' emission and hard X-radiation  (HXR) from some galaxy
clusters represent  new, although less firmly established evidence
for nonthermal activity in galaxy clusters.

The first systematic search for nonthermal X-ray emission from six
galaxy clusters by HEAO-1 resulted only in upper limits
\citep{rephaeli/gruber::HEAO1:1988}.  However, the recent studies with
RXTE and BeppoSAX satellites revealed possible nonthermal X-ray
component from a few clusters of galaxies.  While there is clear
evidence for nonthermal X-rays from Coma cluster
\citep{fusco-femiano/:1999,rephaeli/gruber:2002}, the claims of
presence of nonthermal X-rays in the spectra of other objects, e.g.
Abell 2199 \citep{kaastra/:1999} and Abell 2256
\citep{fusco-femiano/::A2256:2000}, need further confirmation. The
detection of diffuse ``excess'' radiation at lower energies (in the
EUV band) from Coma \citep{lieu/::coma:1996,bowyer/:1999} and Virgo
\citep{Bowyer/::virgo:1996,berghoefer/::virgo:2000} was initially
interpreted as thermal radiation of warm intracluster medium, but soon
it became clear that this radiation has most likely nonthermal origin
\cite[see e.g.][]{berghoefer/bowyer:2002}.  Later, using the data of
BeppoSAX LECs observations, \cite{kaastra/:1999} reported detection of
nonthermal emission from Abell 2199, however
\cite{bowyer/:1999,berghoefer/bowyer:2002} argued that this, as well
as the claims of ``excess'' EUV emission from some other galaxy
clusters (except for Coma and Virgo) were not adequately justified. On
the other hand, the community seems quite confident in detection of
nonthermal high energy radiation from the Coma cluster in both EUV and
X-ray bands.

The radio and EUV/X-ray emission components from galaxy clusters could
be tightly coupled, i.e.  produced by the same population of electrons
through the synchrotron and inverse Compton (IC) channels of
radiation, respectively \citep{rephaeli:1977}.  Since the main target
photons for the inverse Compton scattering are provided by the 2.7 K
CMBR, the ratio of the radio and EUV/HXR fluxes depends significantly
($\propto B^{-2}$) on the intracluster magnetic field (ICMF), but is
rather insensitive to the specific energy distribution of electrons.
Since last several years the inverse Compton origin of the EUV/HXR
``excess'' has been explored by many authors
\citep[e.g.][]{hwang:1997,bowyer/berghoefer:1998,sarazin/lieu:1998,
  atoyan/volk:2000,petrosian:2001, brunetti/:2001,tsay/:2002}.  
The general conclusion of  these studies is that if one
interprets EUV and HXR emission as a result of inverse Compton
scattering of electrons responsible for diffuse radio emission, ICMF
cannot significantly exceed $0.1-0.2\mu$G, unless one introduces quite
unusual assumptions concerning the energy spectrum and the pitch angle
distributions of electrons.  This is an order of
magnitude lower than follows from Faraday Rotation Measures (RM) for
several galaxy clusters \citep{clarke/kronberg/:2001}.  Although, this
discrepancy could be somewhat reduced within more sophisticated IC
models and by adequate treatment of observational selection effects
\citep{petrosian:2001,rephaeli/gruber:2002}, it is important to search
for alternative approaches and mechanisms for explanation of the
EUV/HXR excess.  \cite{atoyan/volk:2000} and \cite{tsay/:2002} have
shown that formally it is possible to accommodate, within the
two-component IC models, the EUV excess from Coma even with high
intracluster magnetic field. At the same time, they demonstrated that
the hard X-ray emission cannot be reproduced by any inverse Compton
scenario if magnetic field exceeds a few $\mu$G.  It is clear that for
explanation of hard X-ray emission from Coma one needs alternative
radiation mechanism(s).

In particular, it has been suggested that HXR might be explained by
bremsstrahlung of supra-thermal electron population
\citep{ensslin/:1999,blasi:2000,dogiel:2000,sarazin/kempner:2000} This
model, however, leads to unreasonably high energy requirements, given
the fact that only a negligible ($\sim 10^{-6}$) fraction of
sub-relativistic electrons is released in the form of bremsstrahlung
X-rays \citep{petrosian:2001}. Recently \cite{dogiel/:2002} suggested
a revised version of this model assuming quasi-thermal distribution of
electrons responsible for X-rays, and argued that this model could
partly avoid above mentioned energetic problem.

In this paper we propose a new scenario for production of nonthermal
EUV and X-ray emission in clusters of galaxies by {\em synchrotron
  radiation} of ultrarelativistic electrons. The key assumption of the
model is that these electrons have ``photonic'' origin, i.e. are
produced at interactions of very high energy (VHE) $\gamma$-rays with
diffuse extragalactic background radiation.  In this way the
relativistic electrons are continuously implemented throughout the
entire intracluster medium, and therefore, the resulting synchrotron
EUV and X-ray emission should have diffuse morphology even in the
case of a single point source of VHE $\gamma$-rays located in the
cluster.

Below we discuss implications of this hypothesis for high energy
nonthermal radiation from clusters of galaxies with some specific
calculations for the Coma cluster  which shows the strongest evidence
for nonthermal emission at EUV and X-ray bands.

\section{The model: motivations and basic assumptions}
   
The proposed scenario is based on two assumptions: (i) the nonthermal high
energy radiation of galaxy clusters is dominated by synchrotron
radiation of ultrarelativistic electrons; (ii) these electrons have
non-acceleration origin, namely,  they  are produced throughout the entire
cluster volume in interactions of hypothetical primary very high
energy $\gamma$-rays with the diffuse extragalactic radiation fields.

The synchrotron radiation of relativistic electrons is so far the most
effective mechanism for production of X-rays in conditions typical for
intracluster medium (ICM).  Indeed, the characteristic time of
radiation of synchrotron X-ray photons with energy $E_{\rm X}$:
$t_{\rm syn} \simeq 5\times 10^4 (E_{\rm X}/1\: \mbox{keV})^{-1/2}
(B/1\: \mu\mbox{G})^{-3/2}$~yr, 
is much shorter than any other
relevant timescale characterizing radiative and other losses of
electrons in these objects. Therefore for production of the EUV/HXR
flux $F_{\rm EUV/X}$ reported from the Coma cluster on the level of a
few times~$10^{-11} \ \rm erg/cm^2 s$ a reasonable injection power of
multi-TeV electrons in ICM is required, 
$\dot{W}_{\rm e} \simeq L_{\rm EUV/X}=4 \pi d^2 \, 
 F_{\rm EUV/X} \sim 3 \times 10^{43} \ \rm erg/s$, 
where $d$ is the distance to the cluster (for Coma $d\sim 100$~Mpc).
Note that this estimation is almost independent on the 
exact value of ICMF as long as the B-field exceeds $3 \mu \rm G$.  The
{\em current} overall energy in these electrons, 
$W_{\rm e} \simeq \dot{W}_{\rm e} t_{\rm syn}$, 
does depend on the magnetic field.  For
$B=3 \mu \rm G$, $W_{\rm e} \sim 10^{55} \ \rm erg$.
Assuming, a broad band (e.g. power-law) injection spectrum of
electrons, this estimate can be increased by one or two orders of
magnitude (depending on the spectral index and the low energy cutoff
in the electron spectrum), but in any case it remains well below the
total energy for electrons with energy between 100 MeV and a few GeV
-- $W_{\rm e} \sim \dot{W}_{\rm e} t_{\rm IC} \sim 10^{62} \ \rm erg$,
required by the IC models.

On the other hand, the high efficiency of synchrotron radiation of
multi-TeV electrons in ICMF leads to other problems.  It is {\em
  over-efficient} in the sense, that the short radiative cooling time
of electrons requires an adequate acceleration rate in order to boost
electrons to energies well beyond 100 TeV.  Note, that in order to
produce synchrotron photons of energy $E_{\rm X}$ one needs electrons
with energy $E_e \simeq 140 (E_{\rm X}/1\: \mbox{keV})^{1/2}
(B/3\:\mu\mbox{G})^{-1/2}$~TeV.  Although very high energy electrons
can be effectively accelerated by strong accretion or merger shocks
\citep{loeb/waxman:2000, inoue/sasaki:2001, miniati:2002,
  blasi:2002,gabici/blasi:2003}, even for most favorable conditions
(allowing the acceleration to proceed in the Bohm diffusion regime),
the maximum electron energy cannot achieve $\geq 100$~TeV which is
required to explain the nonthermal X-ray spectrum of the Coma cluster
reported up to 80 keV.  Also, the short lifetime does not allow
electrons to propagate away from the acceleration sites more than 1
kpc.  Thus, the diffusive character of the observed EUV and HXR cannot
be explained, unless we assume continuous (in space and time)
production of electrons throughout the cluster.  While the direct
acceleration of multi-TeV electrons on $\sim 1$~Mpc scales in the
intracluster medium seems a rather unrealistic scenario, such
energetic electrons can be implemented continuously in the cluster as
{\em secondary products} of interaction of high energy protons and
$\gamma$-rays with ambient matter and photon fields
\citep{aharonian:2002}.

Since protons in galaxy clusters can be accelerated and effectively
confined up to energies $\geq 10^{3} \ \rm TeV$
\citep{voelk/:1996,berezinsky/blasi/:1997} they unavoidably produce
relativistic electrons through generation and decay of secondary
$\pi^\pm$-mesons.  However, because of the low density of the
intra-cluster gas, $n \leq 10^{-3} \ \rm cm^{-3}$, the p-p interaction
time exceeds $10^{19} \ \rm s$, therefore the total energy in
multi-TeV protons should be at least $10^{63} \ \rm erg$ in order to
explain the EUV and HXR fluxes observed from the Coma cluster by
electrons of ``hadronic'' origin.

%
\begin{figure}[t]
\begin{center}
  \includegraphics[width=1.\linewidth]{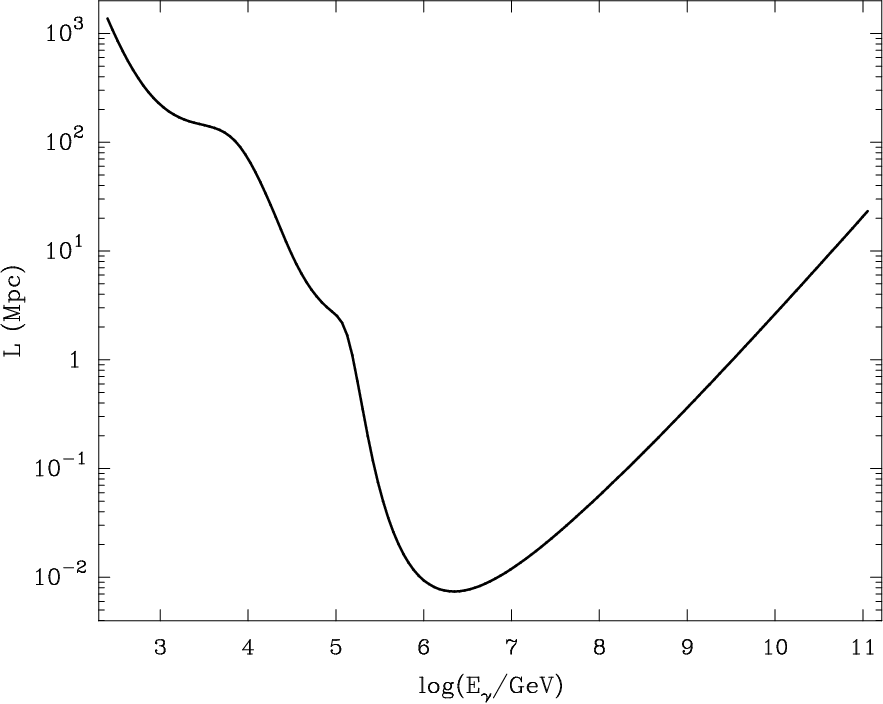}
\caption{ Mean free path of $\gamma$-rays calculated for the diffuse
  extragalactic background model described in \cite{timokhin/:2002} }
\label{fig_mfp}
\end{center}
\end{figure}  
%

The principal assumption of our model is that the ultrarelativistic
electrons responsible for nonthermal high energy emission of galaxy
are products of interaction of very high energy $\gamma$-rays
($10^{14}-10^{19}$~eV) with soft photons of the cosmic diffuse
background radiation. The mean free path of such photons is shown in
Fig.~\ref{fig_mfp}.  It varies in the range from 0.1 to 1~Mpc, which
allows continuous injection of electrons in the entire volume of the
cluster. This overcomes two problems that we face in the case of
directly accelerated electrons -- both related to the short
synchrotron cooling time of electrons.  The hypothesis of secondary
origin of electrons do not have any intrinsic upper limit on the
energy of electrons, and allows implementation of electrons in every
point of ICM.

\section{Results}

The energy of the secondary, pair-produced electrons, and therefore
the typical energy of their synchrotron radiation strongly ($\propto
E^2$) depends on the high energy end of the primary gamma-radiation.
Below we discuss 3 different cases when spectra of primary
$\gamma$-rays extend to (i) ``conventional'' energies of about
$10^{15} \ \rm eV$ or less, (ii) to ultra-high energies exceeding
$10^{16} \ \rm eV$, and (iii) extremely high energies extending to
beyond $10^{20} \ \rm eV$. Although in all three cases the proposed
mechanism works with almost 100 per cent efficiency transforming the
energy of primary $\gamma$-rays to synchrotron radiation of
pair-produced electrons, the resulting synchrotron radiation appears
in essentially different energy bands. We apply this model for
interpretation of the EUV and X-ray ``excess'' emission from the Coma
cluster. Also, we discuss this mechanism in the context of future
searches of $\gamma$-rays from clusters of galaxies.

\subsection{Primary $\gamma$-rays with energy less than 
  $10^{15}\ \rm eV$}

We assume that a source or an ensemble of sources in the central part
of the cluster radiate $\gamma$-rays with a constant rate and energy 
spectrum given by a  ``power-law with quasi-exponential
cutoff'':
\begin{equation} 
\dot{N}_\gamma(E) \propto E^{-\Gamma} 
\exp\left[ ( E/E_0 )^{-\beta} \right]\ .
\label{injspec}
\end{equation} 
We use in calculations the energy-dependent mean free path of
gamma-rays shown in Fig.\ref{fig_mfp}.  We calculate numerically the
distribution of electrons and positrons (injection spectrum) produced
by $\gamma$-rays from the central source(s) in each point of the
cluster.  The time-dependent spectra of pair-produced electrons are
obtained taking into account their synchrotron and IC energy losses.
Finally, spectra of synchrotron and IC radiation of these electrons
are calculated.  We assume that the secondary, pair-produced electrons
are immediately isotropized, We also ignore the propagation effects of
these electrons, i.e, assume that the electrons ``die'', due to severe
synchrotron losses, not far from their birthplace.  For any reasonable
intracluster magnetic fields this is a quite acceptable approximation.
The propagation effect may become relatively important for low-energy
electrons responsible for radio emission.  However, within the assumed
model we do not attempt to explain the radio emission, but rather
assume that radio emission is due to relatively low energy electrons
associated with nonthermal phenomena, that took place long time ago
\citep[see e.g.][]{ensslin/sunyaev:2002}.  At the same time, radio
emission produced by the cooled low energy (MeV/GeV) electrons in the
framework of our model should not exceed the observed fluxes. This is
an important condition which sets robust upper limit on the active
time of operation of $\gamma$-ray sources and on their energy spectra.

\begin{figure}[t]
\begin{center}
\includegraphics[angle=-90,width=1.\linewidth]{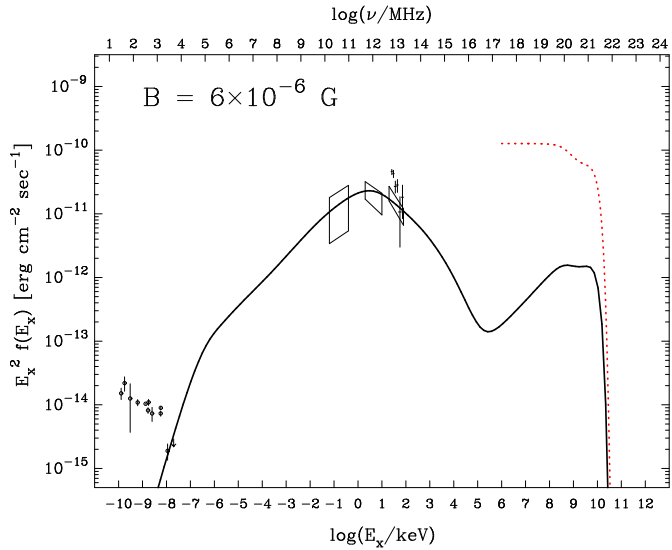}

\caption{Spectrum of non-thermal radiation from the  Coma cluster
  (solid line).  The radiation is produced via conversion of VHE
  $\gamma$-rays into electron-positron pairs and subsequent
  synchrotron and IC radiation of secondary electrons. It is assumed
  that $\gamma$-rays with spectrum given by Eq.(\ref{injspec}) with
  $\Gamma=2$, $\beta=0.5$, $E_0=700$~TeV, have been injected into ICM
  over the last $10^7$ years.  The dotted line represents the spectrum
  of primary $\gamma$-rays as seen by the observer, assuming that
  $\gamma$-rays are emitted isotropically. Both spectra are corrected
  for absorption in the intergalactic medium.  The compilation of
  radio data are taken from \cite{deiss/:1997}.  The open boxes in the
  EUV and soft X-ray domains correspond to the fluxes observed by
  \cite{lieu/:1999}. The open box and points in hard X-ray domain
  correspond to fluxes reported by \cite{fusco-femiano/:1999}. The
  magnetic field in the cluster is assumed to be $B=6\: \mu$G. }
\label{fig_spectr_centr_src_highB}
\end{center}
\end{figure}  

\begin{figure}[t]
\begin{center}

  \includegraphics[angle=-90,width=1.\linewidth]{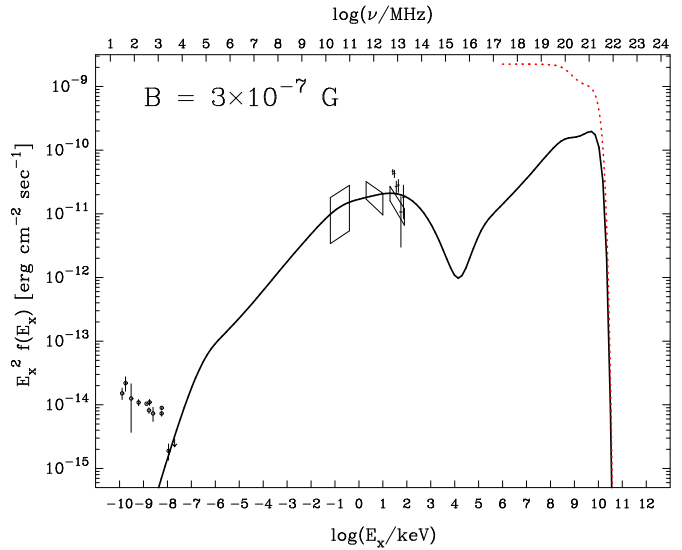}
\caption{The same as in Fig.~\ref{fig_spectr_centr_src_highB}, but for
 the intracluster magnetic field $B=0.3\: \mu$G}
\label{fig_spectr_centr_src_lowB}
\end{center}
\end{figure}  

In Figs.~\ref{fig_spectr_centr_src_highB} and
\ref{fig_spectr_centr_src_lowB} we show the synchrotron and IC spectra
of pair produced electrons in ICM, calculated for two values of the
average ICMF: (a) $B=6\:\mu$G and (b) $B=0.3\: \mu$G, respectively.
For both cases the following parameters of the $\gamma$-ray spectrum
are assumed: $\Gamma=2, \beta=1/2, E_0=700\: \mbox{TeV}$.  It is
assumed that injection of $\gamma$-rays with a quasi-constant rate
into ICM have been started $10^7$~yr ago.  The fluxes shown are
corrected for the intergalactic absorption due to interactions with
the diffuse extragalactic background radiation fields  adopting
the distance to the source of 100 Mpc.

Figs.~\ref{fig_spectr_centr_src_highB} and
\ref{fig_spectr_centr_src_lowB} demonstrate that for the chosen
combination of parameters it is possible to explain the EUV and X-ray
radiation by electrons of ``photonic'' origin.  VHE $\gamma$-ray
luminosities of about $2\times10^{45}$ and $3\times 10^{46}$~erg/s are
requited to support the reported EUV and X-ray fluxes for the cases
(a) and (b) respectively.  Assuming a low energy cutoff in the
$\gamma$-ray spectra below 100 TeV one may significantly reduce these
energy requirements.  In the case (a) the value of ICMF is in
agreement with the estimates derived from RM probes of the Coma
cluster \citep{kim/kronberg/:1990,feretti/::coma_B:1995}.  From the
point of view of energy requirements, the preference obviously should
be given to the case of strong magnetic field too.  Actually, the case
of low magnetic field, which predicts energy flux of the inverse
Compton TeV $\gamma$-rays on the level of $\sim (2-3) \times 10^{-10}
\ \rm erg/cm^2 s$ (see Fig.~\ref{fig_spectr_centr_src_lowB}), is
excluded by TeV observation.  Such a flux hardly could be missed from
the long-term observations of Coma by the HEGRA system of imaging
atmospheric Cherenkov telescopes, even taking into account the
extended character of this emission.  The case of strong magnetic
field predicts significantly reduced secondary $\gamma$-ray flux
(Fig.~\ref{fig_spectr_centr_src_highB}), which however remains
sufficiently high, so it can be probed by the next generation of
Cherenkov telescope arrays.

In addition to the secondary $\gamma$-rays of IC origin, we should
expect also primary $\gamma$-rays, even after intergalactic absorption
which becomes significant already at energies above several TeV.  The
flux of primary $\gamma$-rays, determined by the normalization to
provide the EUV and HXR fluxes by synchrotron radiation of secondary
electrons, is shown in Figs.~\ref{fig_spectr_centr_src_highB}~and%
~\ref{fig_spectr_centr_src_lowB} by dotted line.  It marginally agrees
with the EGRET upper limit at MeV/GeV energies, and exceeds
sensitivity of the HEGRA telescope system even assuming that the
source is extended with angular size of about 1 degree, or
$\gamma$-ray flux is contributed by many sources distributed over the
cluster.  This discrepancy can be removed if we assume anisotropic
$\gamma$-ray emission, e.g.  that $\gamma$-rays are contributed by a
limited number of AGN with jets away from the direction to the
observer.  The conflict with the TeV upper limits can be overcome also
assuming that the primary $\gamma$-ray spectrum is harder or contains
low energy cutoff around 100 TeV.  Such a cutoff could be a result of
$\gamma$-ray absorption in the UV/optical/NIR fields inside the
primary source(s).  While with such an assumption we may dramatically
reduce the flux of primary $\gamma$-rays in the observable ($E \leq 10
\ \rm TeV$) energy domain, it cannot have a significant impact on the
of synchrotron EUV and X-rays, because they are result of interactions
of $E \geq 100 \ \rm TeV$ photons with the diffuse background
radiation.

Meanwhile, spatial profiles of the synchrotron radiation of the
photoproduced electrons significantly depend both on the ICMF
distribution and on the high energy cutoff in the spectrum of primary
$\gamma$-rays.  The dependence on the flux of the diffuse background
radiation is less significant.

\begin{figure}[t]
\begin{center}
  \includegraphics[angle=0,width=1.\linewidth]{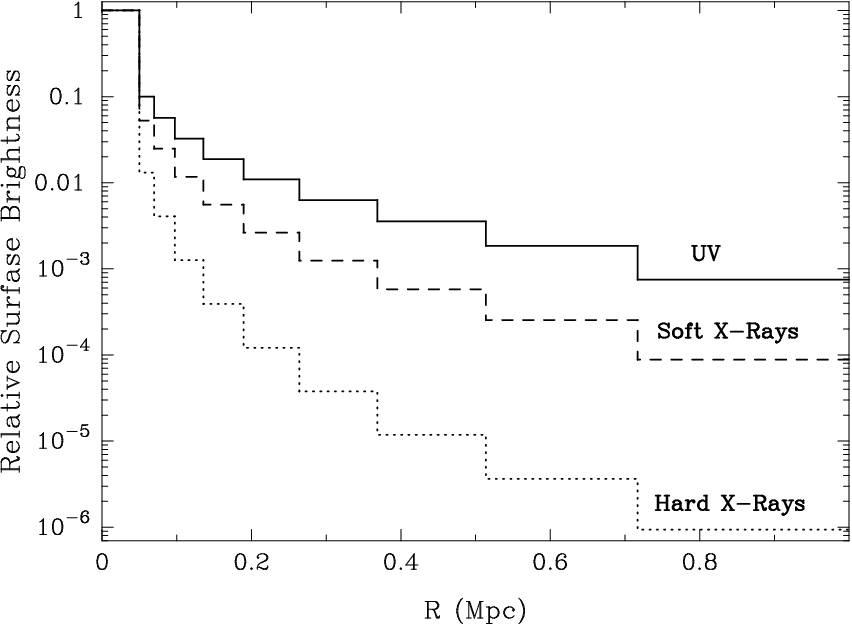}
\caption{Relative surface brightness distribution
  of the non-thermal radiation from Coma cluster in different spectral
  bands for initial $\gamma$-rays produced in central source. The cluster
  is divided in 10 zones. Surface brightness is normalized to the
  brightness of the central zone.  In each zone a separate spectrum
  calculation has been performed. This figure corresponds to the case
  with large ICMF ($B=6\: \mu$G).  Relative surface brightness
  distribution in EUV energy band (0.064 - 0.240 keV) is shown by
  solid line, in Soft X-Rays (2 - 10 keV) by dashed, and in Hard
  X-Rays (25 - 80 keV) by dotted lines.  }
\label{fig_sbright}
\end{center}
\end{figure}

In Fig.~\ref{fig_sbright} we show the surface brightness distribution
of synchrotron radiation in different energy bands, assuming that the
source(s) of primary $\gamma$-rays are concentrated in the central
part of the cluster.  The profiles are calculated for the same
parameters as in Fig.~\ref{fig_spectr_centr_src_highB}. It is seen
that with decrease of photon energy, the brightness distribution
becomes broader. This reflects the reduction of the free path of
$\gamma$-rays with energy.  In this calculations we assume the average
intracluster magnetic field of $6 \ \mu \rm G$. In reality, the
intracluster magnetic field should, of course, decrease at larger
distances from the center.  This should lead to sharper profiles,
especially at high energies.  On the other hand, if sources of primary
$\gamma$-rays are more or less homogeneously distributed in the
cluster, we should expect quite flat brightness distributions at all
photon energies.
 
Comprehensive spectral and morphological studies of high energy
synchrotron components in the spectra of galaxy clusters can provide a
decisive test of the proposed model, and in case of confirmation would
reveal unique information about the energy spectrum and angular
distribution of ``invisible'' (absorbed in the way between the source
and the observer) $\geq 100$~TeV primary $\gamma$-rays.

\subsection{Primary $\gamma$-rays with energy exceeding  
  $10^{16}\ \rm eV $}

The energy of a $\gamma$-ray photon interacting with background
radiation fields is shared between the secondary electron and
positron.  However, the major fraction of the energy of the
$\gamma$-ray photon is transfered to one of the electrons.  Therefore
the maximum of the synchrotron radiation of secondary electrons is
expected at energy $h \nu \sim 10 (B/1 \ \mu \rm G) (E_\gamma/10^{15}
\ \rm eV)^2 \ \rm keV$.  Thus, if the spectrum of $\gamma$-rays
extends beyond $10^{17} \ \rm eV$, the maximum of synchrotron
radiation will be shifted to the $\gamma$-ray domain. Therefore, the
spectrum of synchrotron radiation dramatically depends on the position
of the cutoff energy $E_0$ in the primary $\gamma$-ray spectrum.  For
example, for $E_0=10^{16} \ \rm eV$, we obtain $h \nu \sim 1 \ \rm
MeV$, while for $E_0=10^{18} \ \rm eV$, the synchrotron maximum
appears in the energy range around 10~GeV.

\begin{figure*}
\begin{center}
\includegraphics[width=.98\linewidth]{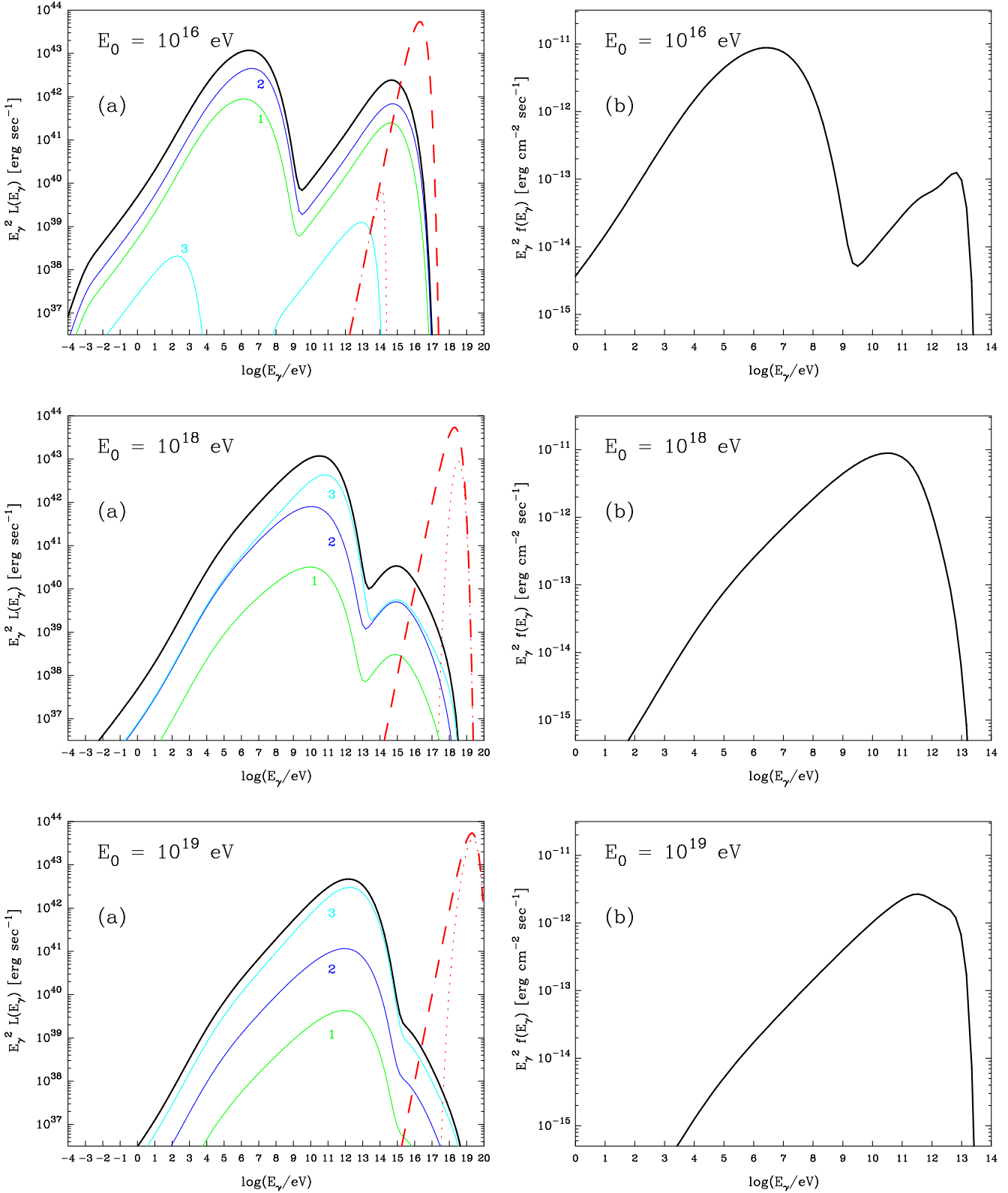}
\caption{%
  {\bf Left panels (a):} Non-thermal luminosities of 
  a   spherically symmetric galaxy cluster of  radius 1.5 Mpc.   
  The radiation is produced via conversion of VHE $\gamma$-rays into
  electron-positron pairs and subsequent synchrotron and IC radiation
  of secondary electrons.  It is assumed that $\gamma$-rays with
  spectrum, given by Eq.(\ref{injspec}) with $\Gamma=0$, $\beta=1$,
  and (i) $E_0=10^{16}$~eV, (ii) $10^{18}$~eV and (iii) $10^{19}$~eV,
  have been injected into ICM by a central source during the last
  $10^7$ years.  The spectra of the primary $\gamma$-ray source are
  shown by thick dashed lines.  Dotted lines show the spectra of the
  primary $\gamma$-rays after propagating of 1.5 Mpc (at the edge of
  the cluster). The area between the dashed and dotted lines indicates
  the amount of energy absorbed and then reradiated in the cluster
  volume. The luminosity of the primary VHE $\gamma$-ray source is
  assumed to be $10^{44}$~erg/sec.  The magnetic field in the cluster
  is assumed to be $B=1\:\mu$G.
  The luminosities of three different zones of the cluster are shown
  with thin solid lines: \textbf{1} ($R=0 - 10^{-3}$~Mpc), \textbf{2}
  ($R=.01 - 0.04$~Mpc) and \textbf{3} ($R=0.44 - 1.5$~Mpc).  The
  overall luminosity of the cluster is shown by thick solid line.
  {\bf Right panels (b):} Expected energy fluxes of non-thermal
  radiation corresponding to the luminosities shown in the right
  panels, calculated for a cluster at a distance of 100 Mpc, after
  correction for the intergalactic absorption.  }
\label{fig_spectr_gamma}
\end{center}
\end{figure*}

This effect is demonstrated in Fig.~\ref{fig_spectr_gamma} assuming
that primary $\gamma$-rays have energy distribution given by
Eq.~(\ref{injspec}) with $\Gamma=0$; $\beta=1$, and (i) $E_0=10^{16} \ 
\rm eV$, (ii) $10^{18} \ \rm eV$ and (iii) $10^{19} \ \rm eV$,
respectively. Note that these type $\gamma$-ray spectra can be formed
by ultrahigh energy cosmic ray protons interacting with narrow-band
radiation field. In this case $E_0$ is an order of magnitude less than
the energy cutoff in the parent proton spectrum. On the other hand,
the low energy part of the $\gamma$-ray spectrum (${\rm d}N/{\rm d}E =
const$) does not depend on the proton spectrum, but simply is result
of the threshold of photomeson interactions.

The $10^{16}-10^{18} \ \rm eV$ $\gamma$-ray can not only be
effectively produced in AGN, but also can escape the production
regions without catastrophic losses \citep[see
e.g.][]{neronov/:2002,atoyan/dermer:2003}.  Another site for
production of extremely high energy gamma- rays could be the
intracluster medium, where the highest energy, $E\geq 10^{20} \ \rm
eV$, cosmic rays interact with the 2.7~K CMBR \citep{aharonian:2002}.
These interactions lead to copious production of secondary
$\gamma$-rays, electrons and neutrinos with characteristic energy
larger than $10^{19} \ \rm eV$.

In Figs.~\ref{fig_spectr_gamma}(a) we show the luminosities of
synchrotron radiation of secondary electrons. It is seen that if in
the case (i) the radiation of the secondary electrons peaks at MeV
energies, in the case (ii) and (iii) the luminosity is dominated by
GeV and TeV $\gamma$-rays, respectively.  However, the very high
energy $\gamma$-rays above 1 TeV suffer significant intergalactic
absorption if sources are located beyond 100 Mpc. To demonstrate this
effect, in Figs.~\ref{fig_spectr_gamma}b we show the expected
$\gamma$-ray fluxes after correction for the intergalactic absorption,
assuming that the source is located at a distance of 100 Mpc.  Thus,
for the assumed total luminosity in primary $\gamma$-rays of $10^{44}
\ \rm erg/s$ and distance to the source of 100~Mpc, the resulting GeV
and TeV $\gamma$-ray fluxes can be probed with GLAST and forthcoming
arrays of atmospheric Cherenkov telescopes. At the same time, because
of limited sensitivity of $\gamma$-ray instruments in the MeV energy
band, detection of the secondary synchrotron radiation initiated by
$10^{16} \ \rm eV$ primary $\gamma$-rays would be very difficult,
unless the power of primary $\gamma$-rays significantly exceeds
$10^{45} \ \rm erg/s$.

\subsection{Non-thermal radiation resulting from decay of
  super-heavy primordial particles}

In this section we discuss production of secondary electrons by
$\gamma$-rays which are not associated with conventional accelerated
processes, but are products of decay or interaction of primordial
massive particles.  These processes constitute the basis of the
so-called ``top-down'' scenarios of production of highest energy
cosmic rays \citep[for a review see e.g.][]{bhattacharjee/sigl:2000}.

\begin{table}[b]
  \caption{Model Parameters}
  \label{tab:X}
  \centering
  \begin{tabular}{lll}
    \hline
    Fragmentation &   $B$  & $m_{\rm X}$\\
    function      &&\\  
    \hline
    1 & $1.8\:\mu$G~~    & $2\times 10^{12}$~GeV \\
    2 & $3.0\:\mu$G~~    & $4\times 10^{12}$~GeV \\
    3 & $2.0\:\mu$G~~    & $6\times 10^{11}$~GeV \\
    4 & $1.8\:\mu$G~~    & $4\times 10^{12}$~GeV \\
    5 & $1.8\:\mu$G~~    & $2\times 10^{13}$~GeV \\
    \hline
  \end{tabular}
\end{table}

Decay or annihilation of weakly interacting massive particle (WIMP) of
the mass $m_{\rm X}$ results in 2 jets of hadrons, with energy $\simeq
m_{\rm X}/2$ each.  The products of hadronic jets are mostly pions --
$\pi^0, \pi^+,\pi^- (\sim 32\% \ \textrm{in each})$ with a small
fraction of energy released in nucleons ($\sim4\%$).  Decays of
$\pi$-mesons result in $\gamma$-rays, electrons and neutrinos with
mean energy of about $10^{20} \ \rm eV$.  While neutrinos freely
propagate through intergalactic medium, $\gamma$-rays cannot penetrate
deeper than 10 Mpc because of interactions with diffuse extragalactic
radio emission. While most of these $\gamma$-rays typically terminate
outside of the cluster, the electrons from decays of charged
$\pi$-mesons immediately radiate their energy through synchrotron
channel in the form of $\gamma$-rays with mean energy $E \simeq 300
(B/3 \ \mu \rm G)(E_{\rm e}/10^{20} \ \rm eV)^2 \ \rm TeV$.  These
$\gamma$-rays interacting with the 2.7~K CMBR and diffuse infrared
background photons lead to a new generation of electrons which
subsequently produce synchrotron EUV and X-rays as described in the
previous section.  It is remarkable that both the expected/predicted
masses of X-particles and the magnetic fields in the galaxy clusters
are in just ``right'' regions to yield, though the 2-step
electron-photon conversions, a significant fraction of the mass of
X-particles in the form of synchrotron EUV and X-radiation.  Detailed
calculations of this radiation require good knowledge of spectral
distributions of secondary particles or the so-called fragmentation
functions. The simplest form corresponding to a very flat spectrum of
particles is approximated as ${\rm d}N_{\rm i}/{\rm d}x \propto
x^{3/2}$, where $x=2E_{\rm i}/m_{\rm X}$ is the dimensionless energy of the
decay product of type i. Besides this ``crude''
approximation (called below as fragmentation function 1) in the
present study we used four other, theoretically better developed
approximations proposed by \cite{hill:1983} (model 2),
\cite{berezinsky/blasi/:1997} (model 3),
\cite{berezinsky/kacherliess:2001} (model 4) and by
\cite{birkel/sarkar:1998} (model 5).

\begin{figure}[tb]
\begin{center}
  \includegraphics[angle=-90,width=1.\linewidth]{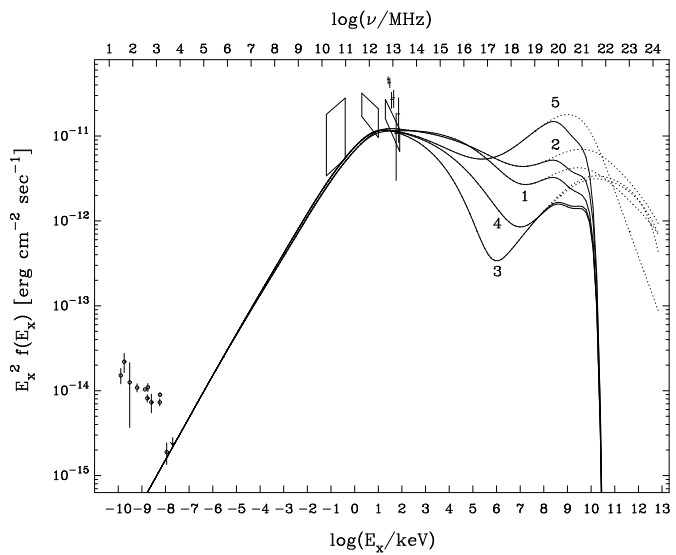}
\caption{Energy fluxes of  non-thermal radiation from the Coma cluster
  initiated by products (electrons, positrons, $\gamma$-rays) of
  decays of super-heavy 'X'-particles.  The spectra are calculated for
  different fragmentation functions, intracluster magnetic field, and
  $m_{\rm X}$ presented in Table \ref{tab:X}.  The reported fluxes at
  radio EUV and X-rays are same as in
  Fig.~\ref{fig_spectr_centr_src_highB}.
  The solid lines represent spectra corrected for intergalactic
  $\gamma$-ray absorption. Unabsorbed spectra are shown by dashed
  lines.  
}
\label{fig_spectr_X}
\end{center}
\end{figure}

We performed calculation for several combinations of the intracluster
magnetic filed strength $B$ and the mass of X-particle $m_{\rm X}$ in
order to find the best fit to the reported fluxes. Results are
presented in Fig.~\ref{fig_spectr_X}.  The model parameters are
presented in Table~\ref{tab:X}.

Since the relic X-particles have cosmological origin, we must assume
that the process continuous over the age of the cluster of about
$1/H_0 \sim 10^{10} \ \rm yr$.  During this period the electrons
cooled down to low energies and thus produce synchrotron radio
emission.  This provides a strict upper limit on the rate of decay of
X-particles in order to prevent overproduction of radio emission.

The results presented in Fig.~\ref{fig_spectr_X} show that with a
normalization at the radio flux at GHz frequencies, and for chosen
values for $B$ and $m_{\rm X}$, all five approximations predict EUV
and X-ray fluxes which marginally agree with observations. However,
the absolute normalizations used in Fig.~\ref{fig_spectr_X} imply very
high rate of appearance of highest energy electrons and $\gamma$-rays.
Namely, the same rate of decay/annihilation of primordial massive
particles in the Halo of our Galaxy, would result in 3 orders of
magnitude higher flux of highest energy particles than the observed
flux of cosmic rays.  In order to avoid overproduction of highest
energy cosmic rays, one has to assume higher density of X particles in
the Coma cluster.  Namely, while in the case of annihilation the
density should be $\sim 30$ times higher than in the Galactic Halo, in
the case of particle decays the density should be $\sim 1000$ times
higher.  Either case can not be easily accommodated within the current
dark matter halo formation theories.

\section{Discussion}

The ``excess'' EUV and hard X-ray emission reported from several
galaxy clusters is generally thought to be due to nonthermal processes
in ICM.  However, interpretation of this radiation within the most
favored inverse Compton model poses serious difficulties.  This model
not only requires huge energy in accelerated particles, but also
assumes intracluster magnetic fields which is significantly below the
values derived from RM probes of galaxy clusters.

The synchrotron radiation of multi-TeV electrons which proceeds with
almost 100 per cent efficiency, seems more viable channel for
production of the ``excess'' EUV and X-ray emission.  The challenge of
this mechanism is the required energy of electrons -- 100 TeV or more.
The lifetime of these electrons and correspondingly their propagation
are very short.  No acceleration mechanism has been suggested so far,
which could overcome the severe radiative losses and boost the
electrons to such high energies. Moreover, such accelerator(s) should
``operate'' on $\gg 1 \ \rm kpc$ scales in order to fill ICM with
multi-TeV electrons.

In this paper we propose that these electrons have secondary origin,
namely we assume that the EUV and X-ray ``excess'' emission is result
of synchrotron radiation of secondary ultrarelativistic electrons
introduced throughout the entire cluster continuously (in time and
space) via interactions of very high energy $\gamma$-rays with photons
from the 2.7 K CMBR and diffuse infrared background radiation.

Of course, generation of extremely high energy photons itself is not a
trivial process, and, in fact, this assumption demands the main
challenging question for the model.  On the other hand, we know that
the $\gamma$-ray spectra of BL Lac objects like Mkn~421 and Mkn~501 do
extend to multi-TeV energies and do not show tendency, after
correction for the intergalactic absorption, for a cutoff at least up
to 20 TeV \citep[see e.g.][]{aharonian/::mkn501newspectr:2001}.
Therefore we may speculate that the energy spectra of these objects
extend to 100 TeV or beyond.  It has been recently argued that
$\gamma$-rays of much higher energies can be produced in more powerful
AGN \citep[e.g.][]{neronov/:2002,atoyan/dermer:2003}, or immediately
in the intracluster medium due to interactions of highest energy
cosmic rays with surrounding photon fields \citep{aharonian:2002}.  In
this case, because the $\gamma$-ray spectra extent well beyond
$10^{15}$~eV, the maximum of synchrotron radiation of the secondary
electrons may appear in the $\gamma$-ray domain.

Also, extremely high energy $\gamma$-rays could be linked to the decay
products of hypothetical super heavy particles from Dark Matter Halos.
In this case the decay products -- $\gamma$-rays and electrons
(positrons) are produced with energies above $10^{21}$~eV. Their
interaction with the surrounding photon and magnetic fields leads to
appearance of the second generation of $\gamma$-rays with energies
typically exceeding 100 TeV.  The latters effectively interact with
the diffuse extragalactic photon fields, resulting in electrons and
positrons of second generation.  And thus, the synchrotron radiation
of these electrons will appear in the X-ray domain.

Above
we applied this scenario for a specific case of Coma -- a relatively
nearby ($d \simeq 100$~Mpc) galaxy cluster showing the strongest
evidence of nonthermal emission in the overall EUV and X-ray bands.
The numerical calculations demonstrate that the reported EUV and X-ray
data can be adequately explained, within experimental uncertainties,
for both high and low intracluster magnetic fields
(see Figs.~\ref{fig_spectr_centr_src_highB}~and%
~\ref{fig_spectr_centr_src_lowB} ).  However the value of the magnetic
field has much stronger impact on the $\gamma$-ray fluxes produced
through the inverse Compton scattering of the same electrons.  In
order to reduce the inverse Compton component of radiation down to
$\sim 10^{-11} \ \rm erg/cm^2 s$ (otherwise a positive signal would be
seen in the HEGRA data) one must assume that the magnetic field
exceeds $3 \:\mu$G.  This agrees very well with the RM probes of Coma
\citep{kim/kronberg/:1990,feretti/::coma_B:1995}.  This also makes
more efficient the production of EUV and X-rays because in this case
all kinetic energy of electrons is released through the synchrotron
channel.

On the other hand, even in the case of high intracluster magnetic
field one may expect quite large fluxes of direct (primary) TeV
$\gamma$-ray emission. In particular, Eq.(\ref{injspec}) with
$\Gamma=2$ predicts quite high direct TeV emission, even after
correction for significant intergalactic absorption.  This flux can
be, however, suppressed assuming that $\gamma$-rays are produced in
jets at large angles towards the observer (this would obviously imply
that $\gamma$-rays are produced by a single or by a small number of
AGN).  Another possibility would be much harder spectrum of primary
$\gamma$-rays (or the existence of a low energy cutoff) below 100 TeV.
While these $\gamma$-rays do not interact effectively with diffuse
background photon within 1 Mpc scales (and therefore they do not have
significant impact on the spectrum of synchrotron radiation), this
assumption can dramatically reduce the flux of primary $\gamma$-rays
at TeV and lower energies.

The production of EUV and X-ray fluxes through synchrotron radiation
requires electrons from 10 to 1000 TeV. Thus, the spectrum of primary
$\gamma$-rays should extent to 1000 TeV or so.  Such energetic
$\gamma$-rays can be produced in individual galaxies, in particular in
AGN, most likely due to proton-proton or proton-photon interactions.
In both cases $\gamma$-ray emission is accompanied with high energy
neutrinos of comparable flux.  In contrast to $\geq 10$ TeV
$\gamma$-rays, which are absorbed in the intergalactic medium, the
neutrinos freely propagate through extragalactic photon fields and
reach us. The expected flux of these energetic neutrinos is expected
at the same level as for $\gamma$-rays if the latters would not be
absorbed in the intergalactic medium, i.e. at the level of $10^{-10} \ 
\rm erg/s$. These fluxes can be detected by planned high energy
neutrino detectors (see e.g. \cite{halzen/:2002}), unless the primary
$\gamma$-rays and, therefore, also neutrinos are not produced in the
jets at large angles to the observer.

The maximum of synchrotron radiation strongly depends on the energy
cutoff in the primary $\gamma$-ray spectrum, $(h \nu)_{\rm max}
\propto E_0^2$.  If $E_0$ significantly exceeds $10^{15}$~eV, the
synchrotron peak will be shifted to the $\gamma$-ray domain, from MeV
to TeV energies, depending on $E_0$.  If so, this scenario can compete
with IC models of $\gamma$-radiation of clusters of galaxies
\citep{waxman/::cluster_gamma_ICS:2003,miniati:2003,gabici/blasi:2003}.
However, while the $\gamma$-ray spectra in the the IC models hardly
could extend beyond 100 GeV (unless particle diffusion in the
unshocked region proceeds in the Bohm regime, which however seems
quite unlikely), in the proposed scenario of synchrotron radiation of
electrons of ``photonic'' origin, the synchrotron peak could be easily
shifted to TeV energies, assuming that the cutoff energy in the
``primary'' $\gamma$-ray spectrum exceeds $10^{18} \ \rm eV$.  The
proposed mechanism works with very high efficiency, providing GeV/TeV
peak luminosities at the level of $10^{43} \ \rm erg/s$ for a
relatively modest luminosity of about $10^{44} \ \rm erg/s$ in the
primary $\geq 10^{18} \ \rm eV$ photons produced in all individual
objects belonging to the cluster. If so, one should expect
non-negligible contribution of $\gamma$-rays from rich clusters of
galaxies into the diffuse extragalactic $\gamma$-ray background.
However, it is difficult to give qualitative estimates of this
contribution given uncertainties in the genuine extragalactic
$\gamma$-ray background due to contamination caused by diffuse
galactic radiation from high latitudes
\citep{strong/moskalenko/:2000}.

In this paper we inspected an interesting
possibility that primary $\gamma$-rays may originate from the decay
products of relic heavy particles like topological defects or WIMPs.
Although  this hypothesis can (marginally) fit the reported
EUV and X-ray spectra from Coma,  it requires strong, by a factor
of 30 to 1000 enhancement of the dark mater density in Coma compared
to the Halo around our Galaxy.   
This makes the hypothesis of cosmological origin 
of primary $\gamma$-rays quite unlikely.

We performed detailed numerical calculation in the framework of the
proposed model for the Coma cluster, which shows so far the strongest
evidence for nonthermal activity in the intracluster medium.  On the
other hand, the mechanism discussed in this paper may have
non-negligible impact on radiation properties of other clusters of
galaxies, especially those, which host strong central AGN or
radiogalaxies, like M87 in the Virgo cluster or NGC~1275 in Perseus.
If one assumes that these objects emit very energetic $\gamma$-rays,
we should expect, in addition to the thermal component of X-rays, also
nonthermal synchrotron radiation produced by secondary, pair produced
electrons. If the energy spectrum of $\gamma$-rays extends to $10^{15}
\ \rm eV$, one may expect significant amount of X-rays from the core
of the cluster, given that the free path of $\sim 10^{15} \ \rm eV$
$\gamma$-rays is only $\sim 10$~kpc.  If so, this hypothetical
component of radiation should be taken into account in the treatment
of X-ray phenomena from clusters of galaxies, in particular in the
context of the on-going debates concerning the ``cooling flows''
\citep[e.g][]{fabian/::missing_soft_X:2002,dar/:2003}.  If identified,
this radiation would also help to ``recover'' information about the
$\geq 100$~TeV $\gamma$-rays which otherwise are not visible due to
the intergalactic absorption. 

Finally we note that search for synchrotron emission of electrons of
``photonic'' origin requires its identification and thorough
separation from the thermal (optically thin bremsstrahlung) component,
taking into account that these radiation components may have rather
similar spectral shapes. Both components can be described by a
power-law spectrum with ``quasi-exponential cutoff''
\begin{equation}
F_\nu\sim \nu^{-\alpha}\exp[(-\nu/\nu_0)^\kappa] \ .
\end{equation}
However, while the spectral index of thermal bremsstrahlung
at energies $h \nu \ll kT$ is $\alpha_{\rm therm}=0$, the spectral
index of the synchrotron component is  close to $\alpha_{\rm
  syn}\approx 0.5$. Thus, for the fixed cutoff energy $h \nu_0$,
detection of an excess emission below $h \nu_0$ may be an indicator of
the synchrotron component.

The cutoff energy for the thermal component is directly related to the
temperature of the intracluster gas, $h \nu_{\rm 0, th} \simeq k T$
(typically a few keV), and the spectrum in the cutoff region drops
exponentially ($k=1$).  The synchrotron cutoff  energy 
depends on the intracluster magnetic field $B$
and on the cutoff energy in the primary photon spectrum $E_0$: 
$h\nu_{0,synch} \simeq 10(B/3 \mu \rm G) (E_0/10^{15} \ \rm TeV)^2$ keV.
For a certain combination of ICMF and the cutoff energy in the
spectrum of primary $\gamma$-rays, the corresponding cutoff energy in the
spectrum of synchrotron radiation may appear around a few keV, i.e.
quite close the the cutoff energy expected in the thermal radiation
component. On the other hand the spectrum of the synchrotron radiation
in the cutoff region is described by the parameter $\kappa \approx
\beta/2$, therefore it drops as a plain exponent only for a specific
distribution of primary $\gamma$-rays given by Eq.(\ref{injspec}) with
$\beta=2$.

Thus, although the position of the cutoff energy in the synchrotron
spectrum may appear in the typical for thermal emission range of about
several keV, the spectrum both below and above the cutoff energy
generally should deviate from the thermal spectrum. It would be a
challenge to search for such a component of radiation from the cores
of galaxy clusters.

\begin{acknowledgements}
  Authors thank V. Petrosian for interesting discussion.  AT
  acknowledges the financial support from INTAS, grant YSF-2002-229,
  and Russian Federation President Grant Program, grants
  NSh-388.2003.2, MK-895.2003.02.
\end{acknowledgements}

\bibliographystyle{aa} 
\bibliography{Xclusters}

\begin{thebibliography}{52}
\expandafter\ifx\csname natexlab\endcsname\relax\def\natexlab#1{#1}\fi

\bibitem[{{Aharonian}(2002)}]{aharonian:2002}
{Aharonian}, F.~A. 2002, \mnras, 332, 215

\bibitem[{{Aharonian} {et~al.}(2001){Aharonian}, {Akhperjanian}, {Barrio},
  {Bernl{\"o}hr}, {Bolz}, {B{\"o}rst}, {Bojahr}, {Contreras}, {Cortina},
  {Denninghoff}, {Fonseca}, {Gonzalez}, {G{\"o}tting}, {Heinzelmann},
  {Hermann}, {Heusler}, {Hofmann}, {Horns}, {Ibarra}, {Iserlohe}, {Jung},
  {Kankanyan}, {Kestel}, {Kettler}, {Kohnle}, {Konopelko}, {Kornmeyer},
  {Kranich}, {Krawczynski}, {Lampeitl}, {Lorenz}, {Lucarelli}, {Magnussen},
  {Mang}, {Meyer}, {Mirzoyan}, {Moralejo}, {Padilla}, {Panter}, {Plaga},
  {Plyasheshnikov}, {Prahl}, {P{\"u}hlhofer}, {Rhode}, {R{\"o}hring}, {Rowell},
  {Sahakian}, {Samorski}, {Schilling}, {Schr{\"o}der}, {Siems}, {Stamm},
  {Tluczykont}, {V{\"o}lk}, {Wiedner}, \&
  {Wittek}}]{aharonian/::mkn501newspectr:2001}
{Aharonian}, F.~A., {Akhperjanian}, A.~G., {Barrio}, J.~A., {et~al.} 2001,
  \aap, 366, 62

\bibitem[{{Aharonian} {et~al.}(2002){Aharonian}, {Timokhin}, \&
  {Plyasheshnikov}}]{timokhin/:2002}
{Aharonian}, F.~A., {Timokhin}, A.~N., \& {Plyasheshnikov}, A.~V. 2002, \aap,
  384, 834

\bibitem[{{Atoyan} \& {Dermer}(2003)}]{atoyan/dermer:2003}
{Atoyan}, A.~M. \& {Dermer}, C.~D. 2003, \apj, 586, 79

\bibitem[{{Atoyan} \& {V{\" o}lk}(2000)}]{atoyan/volk:2000}
{Atoyan}, A.~M. \& {V{\" o}lk}, H.~J. 2000, \apj, 535, 45

\bibitem[{{Berezinsky} \& {Kacherliess}(2001)}]{berezinsky/kacherliess:2001}
{Berezinsky}, V. \& {Kacherliess}, M. 2001, Phys. Rev. D, 63, 034007

\bibitem[{{Berezinsky} {et~al.}(1997){Berezinsky}, {Blasi}, \&
  {Ptuskin}}]{berezinsky/blasi/:1997}
{Berezinsky}, V.~S., {Blasi}, P., \& {Ptuskin}, V.~S. 1997, \apj, 487, 529

\bibitem[{{Bergh{\" o}fer} \& {Bowyer}(2002)}]{berghoefer/bowyer:2002}
{Bergh{\" o}fer}, T.~W. \& {Bowyer}, S. 2002, \apjl, 565, L17

\bibitem[{{Bergh{\" o}fer} {et~al.}(2000){Bergh{\" o}fer}, {Bowyer}, \&
  {Korpela}}]{berghoefer/::virgo:2000}
{Bergh{\" o}fer}, T.~W., {Bowyer}, S., \& {Korpela}, E. 2000, \apj, 535, 615

\bibitem[{{Bhattacharjee} \& {Sigl}(2000)}]{bhattacharjee/sigl:2000}
{Bhattacharjee}, P. \& {Sigl}, G. 2000, \physrep, 327, 109

\bibitem[{{Birkel} \& {Sarkar}(1998)}]{birkel/sarkar:1998}
{Birkel}, M. \& {Sarkar}, S. 1998, Astroparticle Physics, 9, 297

\bibitem[{{Blasi}(2000)}]{blasi:2000}
{Blasi}, P. 2000, \apjl, 532, L9

\bibitem[{{Blasi}(2002)}]{blasi:2002}
---. 2002, Astroparticle Physics, 16, 429

\bibitem[{{Bowyer} \& {Bergh{\" o}fer}(1998)}]{bowyer/berghoefer:1998}
{Bowyer}, S. \& {Bergh{\" o}fer}, T.~W. 1998, \apj, 506, 502

\bibitem[{{Bowyer} {et~al.}(1999){Bowyer}, {Bergh{\" o}fer}, \&
  {Korpela}}]{bowyer/:1999}
{Bowyer}, S., {Bergh{\" o}fer}, T.~W., \& {Korpela}, E.~J. 1999, \apj, 526, 592

\bibitem[{{Bowyer} {et~al.}(1996){Bowyer}, {Lampton}, \&
  {Lieu}}]{Bowyer/::virgo:1996}
{Bowyer}, S., {Lampton}, M., \& {Lieu}, R. 1996, Science, 274, 1338

\bibitem[{{Brunetti} {et~al.}(2001){Brunetti}, {Setti}, {Feretti}, \&
  {Giovannini}}]{brunetti/:2001}
{Brunetti}, G., {Setti}, G., {Feretti}, L., \& {Giovannini}, G. 2001, New
  Astron., 6, 1

\bibitem[{{Clarke} {et~al.}(2001){Clarke}, {Kronberg}, \& {B{\"
  o}hringer}}]{clarke/kronberg/:2001}
{Clarke}, T.~E., {Kronberg}, P.~P., \& {B{\" o}hringer}, H. 2001, \apjl, 547,
  L111

\bibitem[{Colafrancesco {et~al.}(2003)Colafrancesco, Dar, \&
  De~Rujula}]{dar/:2003}
Colafrancesco, S., Dar, A., \& De~Rujula, A. 2003, astro-ph/0304444

\bibitem[{{Deiss} {et~al.}(1997){Deiss}, {Reich}, {Lesch}, \&
  {Wielebinski}}]{deiss/:1997}
{Deiss}, B.~M., {Reich}, W., {Lesch}, H., \& {Wielebinski}, R. 1997, \aap, 321,
  55

\bibitem[{{Dogiel}(2000)}]{dogiel:2000}
{Dogiel}, V.~A. 2000, \aap, 357, 66

\bibitem[{{En{\ss}lin} {et~al.}(1999){En{\ss}lin}, {Lieu}, \&
  {Biermann}}]{ensslin/:1999}
{En{\ss}lin}, T.~A., {Lieu}, R., \& {Biermann}, P.~L. 1999, \aap, 344, 409

\bibitem[{{En{\ss}lin} \& {Sunyaev}(2002)}]{ensslin/sunyaev:2002}
{En{\ss}lin}, T.~A. \& {Sunyaev}, R.~A. 2002, \aap, 383, 423

\bibitem[{{Fabian} {et~al.}(2002){Fabian}, {Allen}, {Crawford}, {Johnstone},
  {Morris}, {Sanders}, \& {Schmidt}}]{fabian/::missing_soft_X:2002}
{Fabian}, A.~C., {Allen}, S.~W., {Crawford}, C.~S., {et~al.} 2002, \mnras, 332,
  L50

\bibitem[{{Feretti} {et~al.}(1995){Feretti}, {Dallacasa}, {Giovannini}, \&
  {Tagliani}}]{feretti/::coma_B:1995}
{Feretti}, L., {Dallacasa}, D., {Giovannini}, G., \& {Tagliani}, A. 1995, \aap,
  302, 680

\bibitem[{{Fusco-Femiano} {et~al.}(2000){Fusco-Femiano}, {Dal Fiume}, {De
  Grandi}, {Feretti}, {Giovannini}, {Grandi}, {Malizia}, {Matt}, \&
  {Molendi}}]{fusco-femiano/::A2256:2000}
{Fusco-Femiano}, R., {Dal Fiume}, D., {De Grandi}, S., {et~al.} 2000, \apjl,
  534, L7

\bibitem[{{Fusco-Femiano} {et~al.}(1999){Fusco-Femiano}, {Dal Fuime},
  {Feretti}, {Giovannini}, {Grandi}, {Matt}, {Molendi}, \&
  {Santangelo}}]{fusco-femiano/:1999}
{Fusco-Femiano}, R., {Dal Fuime}, D., {Feretti}, L., {et~al.} 1999, \apjl, 513,
  L21

\bibitem[{Gabici \& Blasi(2003)}]{gabici/blasi:2003}
Gabici, S. \& Blasi, P. 2003, astro-ph/0306369

\bibitem[{{Giovannini} {et~al.}(1999){Giovannini}, {Tordi}, \&
  {Feretti}}]{giovannini/::radio_clusters:1999}
{Giovannini}, G., {Tordi}, M., \& {Feretti}, L. 1999, New Astronomy, 4, 141

\bibitem[{{Halzen} \& {Hooper}(2002)}]{halzen/:2002}
{Halzen}, F. \& {Hooper}, D. 2002, Reports of Progress in Physics, 65, 1025

\bibitem[{{Hill}(1983)}]{hill:1983}
{Hill}, C.~T. 1983, Nucl. Phys. B, 224, 469

\bibitem[{{Hwang}(1997)}]{hwang:1997}
{Hwang}, C. 1997, Science, 278, 1917

\bibitem[{{Inoue} \& {Sasaki}(2001)}]{inoue/sasaki:2001}
{Inoue}, S. \& {Sasaki}, S. 2001, \apj, 562, 618

\bibitem[{{Kaastra} {et~al.}(1999){Kaastra}, {Lieu}, {Mittaz}, {Bleeker},
  {Mewe}, {Colafrancesco}, \& {Lockman}}]{kaastra/:1999}
{Kaastra}, J.~S., {Lieu}, R., {Mittaz}, J.~P.~D., {et~al.} 1999, \apjl, 519,
  L119

\bibitem[{{Keshet} {et~al.}(2003){Keshet}, {Waxman}, {Loeb}, {Springel}, \&
  {Hernquist}}]{waxman/::cluster_gamma_ICS:2003}
{Keshet}, U., {Waxman}, E., {Loeb}, A., {Springel}, V., \& {Hernquist}, L.
  2003, \apj, 585, 128

\bibitem[{{Kim} {et~al.}(1990){Kim}, {Kronberg}, {Dewdney}, \&
  {Landecker}}]{kim/kronberg/:1990}
{Kim}, K.-T., {Kronberg}, P.~P., {Dewdney}, P.~E., \& {Landecker}, T.~L. 1990,
  \apj, 355, 29

\bibitem[{{Liang} {et~al.}(2002){Liang}, {Dogiel}, \&
  {Birkinshaw}}]{dogiel/:2002}
{Liang}, H., {Dogiel}, V.~A., \& {Birkinshaw}, M. 2002, \mnras, 337, 567

\bibitem[{{Lieu} {et~al.}(1999){Lieu}, {Ip}, {Axford}, \& M.}]{lieu/:1999}
{Lieu}, R., {Ip}, W.-H., {Axford}, W.~I., \& M., B. 1999, \apjl, 510, L25

\bibitem[{{Lieu} {et~al.}(1996){Lieu}, {Mittaz}, {Bowyer}, {Breen}, {Lockman},
  {Murphy}, \& {Hwang}}]{lieu/::coma:1996}
{Lieu}, R., {Mittaz}, J.~P.~D., {Bowyer}, S., {et~al.} 1996, Science, 274, 1335

\bibitem[{{Loeb} \& {Waxman}(2000)}]{loeb/waxman:2000}
{Loeb}, A. \& {Waxman}, E. 2000, \nat, 405, 156

\bibitem[{{Miniati}(2002)}]{miniati:2002}
{Miniati}, F. 2002, \mnras, 337, 199

\bibitem[{{Miniati}(2003)}]{miniati:2003}
---. 2003, \mnras, 342, 1009

\bibitem[{{Neronov} {et~al.}(2002){Neronov}, {Semikoz}, {Aharonian}, \&
  {Kalashev}}]{neronov/:2002}
{Neronov}, A., {Semikoz}, D., {Aharonian}, F., \& {Kalashev}, O. 2002, Physical
  Review Letters, 89, 51101

\bibitem[{{Petrosian}(2001)}]{petrosian:2001}
{Petrosian}, V.~. 2001, \apj, 557, 560

\bibitem[{{Rephaeli}(1977)}]{rephaeli:1977}
{Rephaeli}, Y. 1977, \apj, 212, 608

\bibitem[{{Rephaeli} \& {Gruber}(2002)}]{rephaeli/gruber:2002}
{Rephaeli}, Y. \& {Gruber}, D. 2002, \apj, 579, 587

\bibitem[{{Rephaeli} \& {Gruber}(1988)}]{rephaeli/gruber::HEAO1:1988}
{Rephaeli}, Y. \& {Gruber}, D.~E. 1988, \apj, 333, 133

\bibitem[{{Sarazin} \& {Kempner}(2000)}]{sarazin/kempner:2000}
{Sarazin}, C.~L. \& {Kempner}, J.~C. 2000, \apj, 533, 73

\bibitem[{{Sarazin} \& {Lieu}(1998)}]{sarazin/lieu:1998}
{Sarazin}, C.~L. \& {Lieu}, R. 1998, \apjl, 494, L177

\bibitem[{{Strong} {et~al.}(2000){Strong}, {Moskalenko}, \&
  {Reimer}}]{strong/moskalenko/:2000}
{Strong}, A.~W., {Moskalenko}, I.~V., \& {Reimer}, O. 2000, \apj, 537, 763

\bibitem[{{Tsay} {et~al.}(2002){Tsay}, {Hwang}, \& {Bowyer}}]{tsay/:2002}
{Tsay}, M.~Y., {Hwang}, C., \& {Bowyer}, S. 2002, \apj, 566, 794

\bibitem[{{Volk} {et~al.}(1996){Volk}, {Aharonian}, \&
  {Breitschwerdt}}]{voelk/:1996}
{Volk}, H.~J., {Aharonian}, F.~A., \& {Breitschwerdt}, D. 1996, Space Science
  Reviews, 75, 279

\end{thebibliography}

\end{document}